# Effect of layer number and layer stacking registry on the formation and quantification of defects in graphene

Sara D. Costa, Johan Ek Weis, Otakar Frank, Martin Kalbac[*]

J. Heyrovský Institute of Physical Chemistry, Academy of Sciences of the Czech Republic, v.v.i., Dolejškova 3, CZ-18223 Prague 8, Czech Republic.

Correct defect quantification in graphene samples is crucial both for fundamental and applied research. Raman spectroscopy represents the most widely used tool to identify defects in graphene. However, despite its extreme importance the relation between the Raman features and the amount of defects in multilayered graphene samples has not been experimentally verified. In this study we intentionally created defects in single layer graphene, turbostratic bilayer graphene and Bernal stacked bilayer graphene by oxygen plasma. By employing isotopic labelling, our study reveals substantial differences of the effects of plasma treatment on individual layers in bilayer graphene with different stacking orders. In addition Raman spectroscopy evidences scattering of phonons in the bottom layer by defects in the top layer for Bernal-stacked samples, which can in general lead to overestimation of the number of defects by as much as a factor of two.

1. Introduction

The mechanical and electronic properties of perfect graphene layers are well described; however, defects are commonly found in real graphene samples. As defects are highly reactive sites, they can be useful for sensing applications [1, 2], and for the functionalisation of graphene [3]. Consequently, an increase of the number of defects in graphene can be required. However, the creation of new defects should be done in a controlled way. Disorder in graphene can be categorised into two regimes: a low-defect-density regime, in which $sp^2$-bonded sites are converted into $sp^3$ bonding, and a high-defect-density regime, in which vacancies in the carbon network are created,

---

[*]Corresponding author. Tel: (+420) 266 053 804. Email: martin.kalbac@jh-inst.cas.cz (Martin Kalbac)

eventually leading to the amorphisation of the material [4]. Several techniques can be employed to induce defects in graphene layers, such as ion bombardment [5, 6], laser irradiation [7], or oxygen plasmas [8-10]. In particular, the oxygen plasma seems to be a favourable approach because oxygen plasma instruments are commonly found in laboratories and are used to clean substrates, cure polymers or etch carbon layers. In addition, oxygen plasma treatments represent an easy, and potentially controllable, way to induce defects in graphene layers, as a set of parameters such as power, pressure and time of exposure can be varied. Oxidation of graphene via oxygen plasma treatment has also been shown to be a practical way to achieve enhancement of the Raman signal [9] and to decrease graphene's hydrophobicity [11].

However, an important question is if the oxygen plasma treatment affects graphene monolayers and bilayers (Bernal stacked or not) differently. Bilayer graphene has been widely investigated as it has great potential to be used in field-effect transistors (FETs) because of a better control of the band-gap. McEvoy et al. [10], reported on oxygen plasma treatments applied to single layers of graphene and mentioned the presence of secondary islands, probably bilayers, that were more resistant to the creation of defects than the monolayer background. Nourbakhsh et al. [12], exposed Bernal-stacked bilayer graphene to an oxygen plasma treatment and observed that oxidised 2-LG retains its semimetallic nature, in opposition to defected 1-LG, but after strong plasma treatment the Raman spectra of the defected bilayers became similar to 1-LG. Investigation of the randomly stacked bilayers is equally important, because most chemical vapour deposition (CVD)-grown graphene is a mixture of bilayer grains stacked with different rotational angles [13]. In general, two Bernal-stacked (2-LG AB) layers behave alike, whereas turbostratic (2-LG T) layers can show a more individual behaviour. Although both top and bottom layers in Bernal-stacked graphene show a similar behaviour in the Raman spectrum, it is not clear whether their properties are in fact similar, or if it is an effect of phonon scattering. The stacking order of two graphene layers is known to influence the effective Young's modulus [14], the behaviour of the graphene at high temperatures [15] and the functionalization of graphene, e.g. fluorination [16]. If plasma treatments affect 2-LG AB and 2-LG T bilayers differently, a more detailed study is essential.

In this work, we report on the investigation of oxygen-plasma-induced defects on bilayers of graphene, both turbostratic and Bernal stacked, and of monolayers as well, for comparison. The creation of defects on the top and bottom layers was also investigated, using graphene bilayers con-taining carbon isotopes ($^{13}$C in the bottom layer and $^{12}$C in the top layer). High-resolution Raman mapping was used to obtain large data sets and to improve the accuracy of the results. The defects creation was found to be indeed dependent on the number of graphene layer and their stacking order. In addition we found that the relation between the Raman signatures of defects and the actual amount of defects is also significantly influenced by the presence of another graphene layer and by the stacking order of the graphene layers. This finding is crucial for correct quantification of the number of defects in multi-layered graphene samples based on the Raman spectroscopy data.

2. Experimental

The graphene layers were grown by CVD on Cu foil. Briefly, the Cu foil (1600 mm$^2$, 127 μm thick, 99.9%, Alfa Aesar) used in the growth was placed in the centre of a furnace, heated to 1273 K and then annealed for 20 min under flowing H$_2$ (99.9999 %, Messer), 50 standard cubic centimetres per minute (sccm). 5 sccm of CH$_4$ (99.9995 %, Messer) was then introduced for another 30 min. Afterwards, the samples were annealed in H$_2$ for 5 min and then cooled from 1273 to 273 K, under a H$_2$ flow. The pressure was kept at 0.35 Torr during the whole process. The isotopically labelled graphene samples were obtained following a similar procedure, using $^{12}$C CH$_4$ first and then $^{13}$C CH$_4$ (99.5 atom %, Sigma-Aldrich), as described in reference [17]. In brief, Cu foil used in the growth was placed in the centre of a furnace, heated to 1273 K and then annealed for 20 min under flowing H$_2$ (50 standard cubic sccm), than $^{12}$C CH$_4$ was introduced first for 90 s, to grow a continuous single layer of graphene. Afterwards, the $^{12}$C CH$_4$ flow was stopped and $^{13}$C CH$_4$ was introduced for 30 min. Subsequently, the graphene was transferred onto SiO$_2$/Si substrates via poly(methyl methacrylate) transfer, as described elsewhere [18]. The graphene samples were then placed inside the chamber of the oxygen plasma etcher (PICO, Diener Plasma-Surface Technology, Germany). To find the optimal parameters, the plasma treatment was repeated several times using different pressures and plasma powers. After determining the most suitable parameters, the procedure was as follows. The chamber was first pumped out and then filled with O$_2$. When the set pressure was reached (0.80 mbar), the plasma was turned on (200 W), for 4 s. Afterwards, the plasma chamber was purged. The procedure was repeated several times on the same sample, until an 88 s total time of plasma exposure was reached. Between each plasma treatment, thus every 4 s of plasma treatment, Raman maps were acquired. The Raman maps were acquired using a WiTec Alpha300, equipped with a 532 nm laser and a 50× objective. Each Raman map contains 625 spectra. To ensure an accurate comparison between the different maps, the analysed areas were always identical. The fits of the Raman features were performed using Lorentzian lines.

## 3. Results and Discussion

### 3.1. Effects of plasma exposure on the creation of defects

A plasma originates in a partially ionized gas in an electrically quasi-neutral state. When an electrical current is passed through a gas, the gas breaks down to form a plasma, which is ionized, interacting with any surface exposed to it. Thereby, subjecting a graphene layer to an oxygen plasma treatment, even for times as short as 1 s, introduces oxygenated groups to the carbon lattice, affecting its electronic structure. Strong oxygen plasma treatments (high plasma powers, e.g. 1000 W, or long exposure times) are known to have adverse effects on graphene's mechanical and electronic properties [4, 19]. Therefore, in the study reported here, a low plasma power (200 W) has been used. To confirm that no visible damage is inflicted in the graphene layers, the samples were observed with an optical microscope, before and after plasma treatment, and no visible tearing or damaging was observed (see Figure S2 in the supporting information). The effects of oxygen plasma treatments can, however, easily be followed via Raman spectroscopy, as the introduction of functional groups generates modifications in the $sp^2$ lattice of graphene, i.e. defects. As the disorder increases, the Raman spectrum of graphene evolves into a more intense D band, higher D/G area ratios, and broader peaks in general [20]. Figure 1 presents Raman spectra of 1-LG, 2-LG AB and 2-LG T graphene before (0 s) and after 88 s of plasma treatment obtained from the same area of the sample. For as-grown graphene, before the plasma treatment, the G band is found in the Raman spectra at ≈1590 $cm^{-1}$, while the 2D band is observed at ≈2690 $cm^{-1}$ (measured at 2.33 eV). In addition, the D band, found at ≈1345 $cm^{-1}$, is a defect-dependent Raman feature [20], which increases after the sample is exposed to the plasma. Moreover, for defected samples, a low-intensity band, the D′ band, can be observed at about 1620 $cm^{-1}$. For highly defected graphene layers, the D′ band may merge with the G band, resulting in a single broader band around 1600 cm$^{-1}$. For mild plasma treatments such as those one reported here, the D′ band is only detectable at longer exposure times.

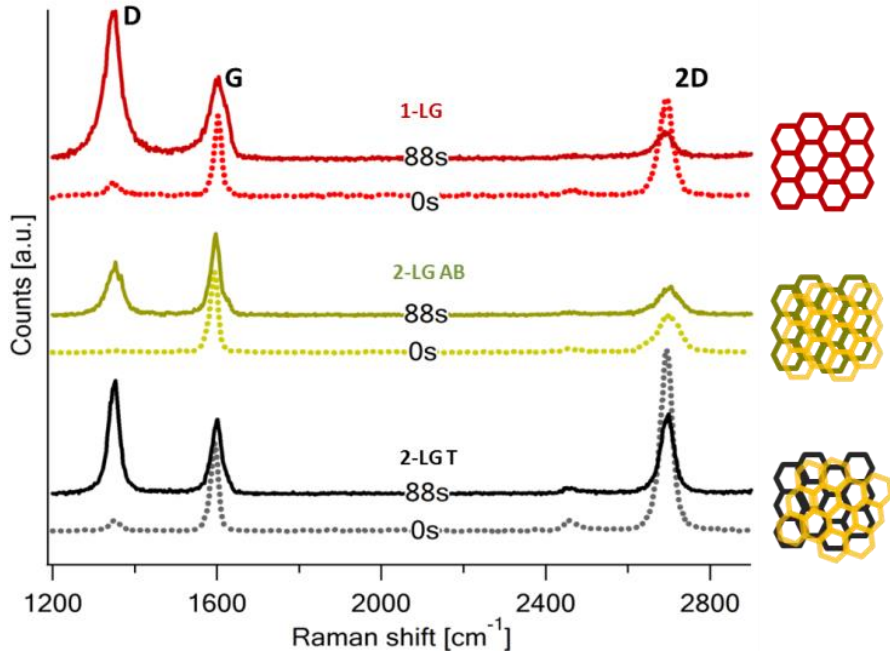

**Figure 1- Raman spectra depicting the main features (D, G and 2D bands) of 1-LG, 2-LG AB and 2-LG T, before and after oxygen plasma treatment.**

Figures 2a and 2b show the optical image and the 2D band full width at half maximum (FWHM) Raman map, respectively, of the analysed area in the graphene sample. Using a single sample containing 1-LG, 2-LG AB and 2-LG T is highly advantageous because it guarantees a more even treatment to all areas. However, it is also necessary to analyse the different areas separately. The 2D band-width varies according to the number of layers and the stacking order of graphene [20]. It is therefore a useful parameter to separate monolayer (30 cm$^{-1}$ ≤ 2D$_{FWHM}$ ≤ 60 cm$^{-1}$ for transferred CVD-grown graphene), bilayer AB (2D$_{FWHM}$ ≥ 60 cm$^{-1}$) and bilayer T (2D$_{FWHM}$ ≤ 30 cm$^{-1}$) areas in the maps, as shown in Figure 2c. In the case of 2-LG T, the specific twist angles lead to formation of the van Hove singularities in the electronic structure of bilayer graphene [21, 22]. If the laser excitation energy matches the energy split between the van Hove singularities, a strong intensity enhancement, and variations in the frequency and FWHM of some Raman modes are observed. For the graphene regions investigated here, the enhancement of the Raman signal is not observed. Thus, it is assumed that the grains under study are not in resonance with the laser energy and the Raman modes are not affected significantly.

In addition variation of the angle between the layers can actually influence the interaction between the top and bottom layer in 2-LG T. These differences in interaction can rationalize small variations in the widths and intensities of the D and 2D modes within the turbostratic areas. Nevertheless, the variations are relatively small and they are all averaged in the statistical evaluation of the measured identical Raman maps.

Figure 2d–f shows the D/G band and D/2D band area ratios for 1-LG, 2-LG AB and 2-LG T, respectively. As the samples get more exposure to the plasma treatment, the values for both D/G and D/2D ratios increase. The behaviour of the D/G and D/2D ratios as exposure to the plasma increases can be attributed to a gradual evolution from $sp^2$-bonded carbon to defected graphene containing $sp^3$ bonding, and eventually reaching an amorphous state [8, 23]. This evolution has been investigated before and is also related to the D band-width behaviour, and will be discussed further in this section. Simultaneously, as the number of defects increases the 2D band area decreases, as observed in Figures 2d–f. The gradual decrease of the 2D band is consistent with previous work [4, 8, 11, 24], and it is assigned to a combination of the *p*-type doping of graphene and the defect-induced suppression of the lattice vibration mode corresponding to the 2D peak [8].

Following the values of D/G and D/2D ratios with the exposure time, in Figures 2d–f, some trends are observed. For 2-LG AB, both D/G and D/2D follow a similar curve, while for 1-LG and 2-LG T the D/G ratio generally has higher values than the D/2D ratio. At exposure times higher than 80 s, for 1-LG graphene, and at 88 s for 2-LG T, the D/2D ratio becomes higher than the D/G ratio, reflecting a larger decrease of the 2D band area compared with the increase of the D band area. The values also differ depending on whether the sample is monolayer, bilayer AB or bilayer T, suggesting that even though the plasma treatment was the same in all areas, different effects are observed for bilayer and monolayer graphene.

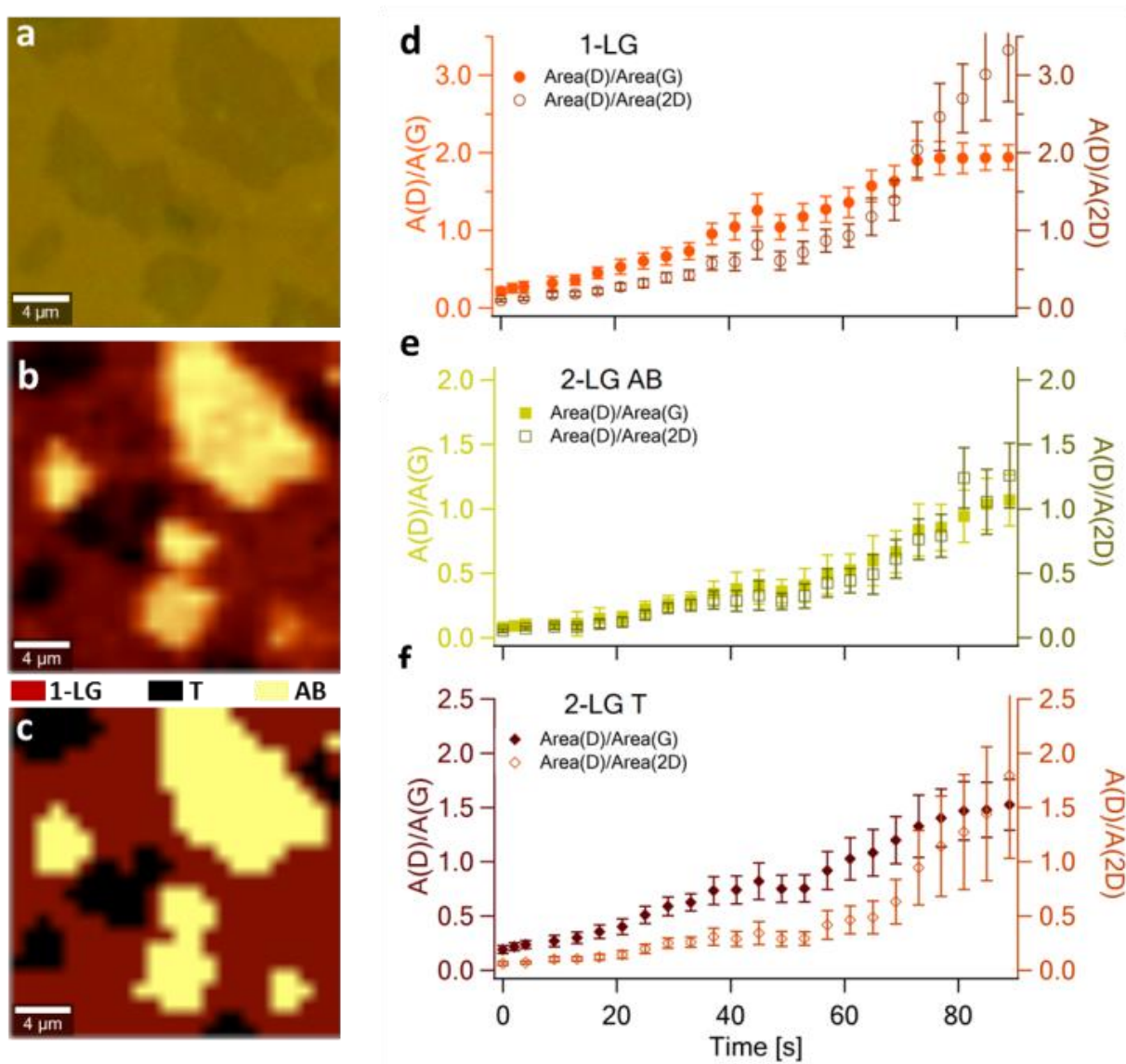

**Figure 2-** a) Optical image of the analysed area; b) Raman map of the 2D band width, showing monolayer (1-LG) together with bilayer turbostratic (2-LG T) and bilayer Bernal (2-LG AB) stacked graphene areas. The 2D band-width is then used to separate spectra according to number of layers, one or two, and their stacking order, AB or T; c) Map highlighting the different regions of the sample: red for 1-LG, black for T 2-LG and yellow for AB 2-LG; Area(D)/Area(G) and Area(D)/Area(2D) ratios plotted as a function of time of oxygen plasma treatment, for d) 1-LG, e) 2-LG AB and f) 2-LG T. Each point in the graphs was calculated using the average of hundreds of Raman spectra, and the error bars represent the standard deviation.

In general, the creation of defects in graphene can be divided into a low-defect-density regime mainly with sp$^3$ defects, in which the D/G ratio increases with disorder, and a high-defect-density regime characterised by the decrease of the D/G intensity ratio with the increase of defects due to the formation of vacancies [4]. The different regimes observed during the creation of defects in a graphene layer are also reflected in the D band line shape. When a small number of defects is induced, the D band increases in amplitude, while for high numbers of defects the D band also broadens, significantly increasing its area [5].

In Figure 3, the D band FWHM is plotted against time of plasma treatment. In as-grown graphene, before any plasma treatment, only a small D band was observed (see Figure 1), probably due to the edges of the grains and point defects. The evolution of the D band FWHM during the plasma treatment can be divided into three stages: I) accentuated decrease (purple area); II) constant and/or soft increase (orange area); III) accentuated increase (pink area). The insets in the figure show spectra of the D band in the different areas, and the arrows highlight that the narrower D band is found in area II. In previous studies involving the creation of defects, a slow increase of phase II and a more abrupt increase of phase III have been observed [25]. The low defect density of the graphene samples is not typically analysed in the literature, hence the initial decrease of the D band FWHM is not indicated.

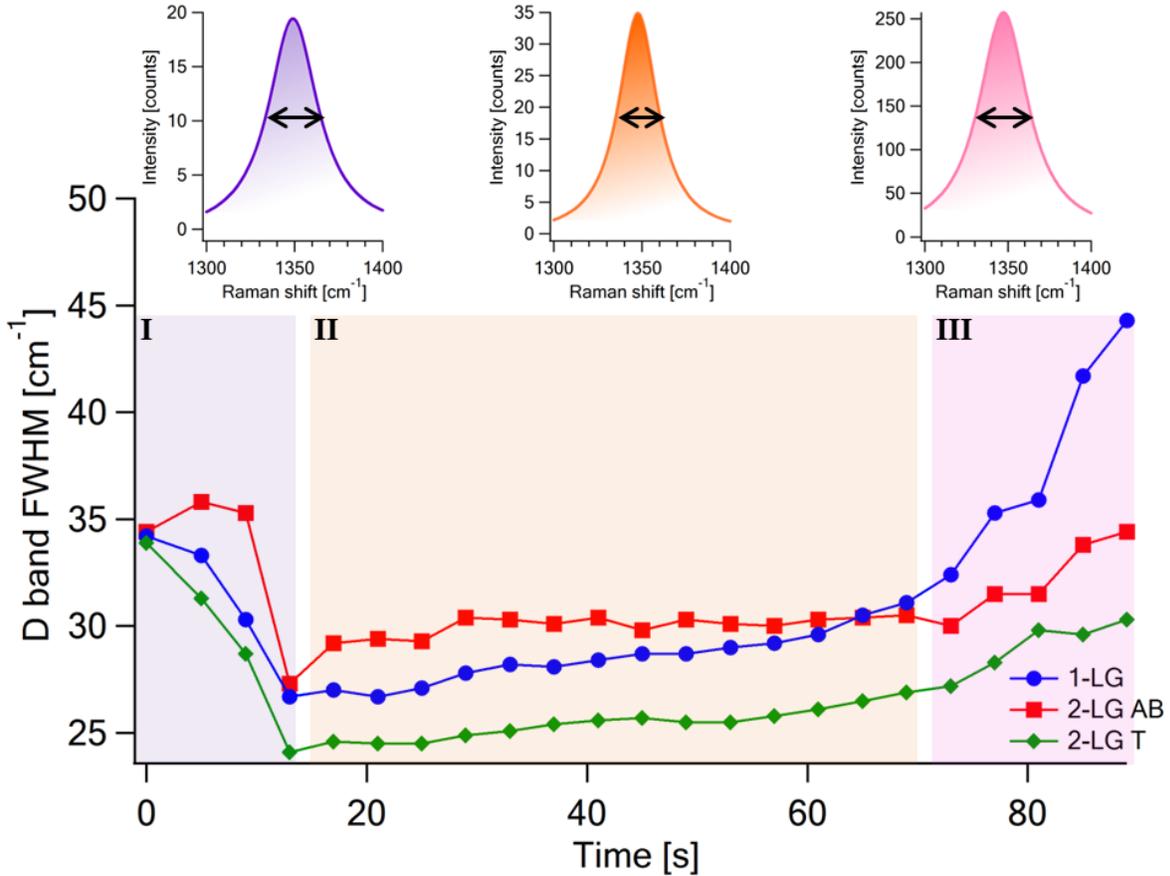

**Figure 3-** D band FWHM plotted as a function of time for 1-LG, 2-LG AB and 2-LG T. The highlighted regions are assigned to the behaviour of the D band FWHM with time of plasma treatment: accentuated decrease (purple), stabilisation and gentle increase (orange), and accentuated increase (pink). Each point in the graphs was calculated using the average of hundreds of Raman spectra. The plots are normalised to the D band FWHM at 0 s. The insets show D band spectra in the different highlighted areas. The black arrows highlight the FWHM of the D band.

The detailed mechanism of defect formation is still unclear due to the complexity of its investigation, since the D band represents many types of defects. Different types of defects can originate D band contributions with slightly different frequencies [26]. It can be hypothesized that in the early stage of oxygen plasma treatment, the different types of defects are transformed into the most stable form, which leads to the narrowing of the D mode. Simultaneously, new random defects are created, which leads to an increase of the D band intensity and FWHM. Both processes com-pete and result in formation of a local minimum in the FWHM as shown in Figure 3. The phase that leads up to the minimum in FWHM was named stage I, while the gradual increase of the D band width and intensity observed afterwards, was named stage II. For treatment times above 68 s, a rapid increase in the D band width is observed, which is most likely associated with alteration to a high-defect-density regime. This effect is more evident for 1-LG than for 2-LG, which is in agreement with the results of Figure 2d, where a rapid increase of the ratio Area(D)/Area(2D) after 68 s is also observed. The intensity ratio between the defect-dependent bands is a good approach to discriminate between $sp^3$-type defects and vacancies: I(D)/I(D′) > 7 indicates the presence of $sp^3$-type defects, assigned to the low-defect-density regime, while I(D)/I(D′) < 7 is assigned to vacancy-type defects, the high-defect-density regime [4, 26]. In our study, the D′ band is observed for 1-LG exposed to the plasma for 56 s or longer and the I(D)/I(D′) ratio is always higher than 7 (see Figure S3 in supporting information), indicating a low-defect-density regime. However, at around t = 68 s, the I(D)/I(D′) ratio drops drastically, from ca. 110 to 50, and keeps decreasing gradually afterwards, reaching a value of 20. For longer exposure times it is possible that carbon atoms have been removed from the carbon network creating vacancies in the graphene layers. Although the I(D)/I(D′) ratio never reaches the low/high-defect-density threshold, its low values indicate that $sp^3$-type defects and vacancies coexist, keeping in mind that each value is obtained from the average of hundreds of Raman spectra.

3.2. Effects of plasma exposure on doping and strain in graphene

A common effect closely related to the creation of defects in graphene is doping. The doping can influence the intensity of Raman features of graphene and consequently the evaluation of the number of defects in graphene [27]. Figure 4 shows the relation between the 2D and G bands frequencies, which can indicate the presence of doping (slope of –0.1 for *n*-type or 0.4 for *p*-type [28]) or strain (slope of ≈2.2 [29]). For 2-LG AB and 2-LG T, the relation between the G and 2D frequencies is linear and has a slope of 0.51 and 0.47, respectively, indicating *p*-type doping. This is in agreement with previous studies of graphene on $SiO_2$/Si substrates exposed to oxygen plasma where *p*-type doping was also detected [9, 12]. Regarding 1-LG, the presence of two different linear relations can be observed: a slope of 0.57 for exposure times below 44 s, and a slope of 1.74 for times above 44 s. The first can be assigned to *p*-type doping, similar to 2-LG, while the second slope points to a mixture of strain and doping. As previously observed, 1-LG enters a high-defect-density regime at times above 68 s, and a different type of defect might occur (e.g. vacancies might start appearing, coexisting with $sp^3$ defects). In this way, as the graphene structure is modified, the relation between the G and 2D frequencies becomes harder to analyse and to assign to doping or strain.

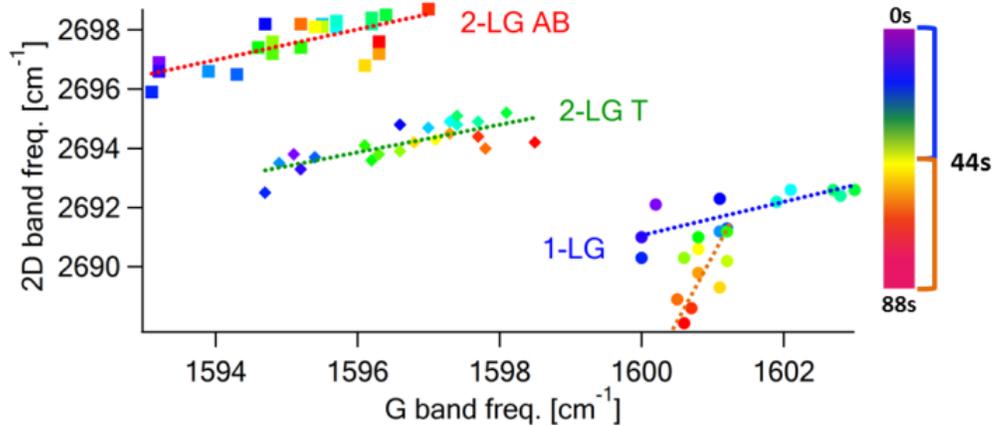

**Figure 4- Correlation of 2D and G bands frequencies for 1-LG, 2-LG AB and 2-LG T. The dashed lines represent the fits for the data. The data obtained for 2-LG AB and 2-LG T increase linearly, while for 1-LG two phases are observed. The time of plasma treatment is represented by the colour scale on the right. Each point in the graphs was calculated using the average of hundreds of Raman spectra.**

3.3. Evaluation of the number of defects in individual layers of graphene bilayers

In graphene bilayer, the Raman signal results from a contribution two carbon layers, hence under normal conditions the individual graphene layers cannot be distinguished. However, individual graphene layers can be isotopically labelled using different carbon isotopes, for example $^{13}C$ on the bottom layer and $^{12}C$ on the top layer. In isotopically labelled samples, the individual graphene layers can easily be addressed as we have shown previously [17, 30-32]. Figure 5 shows the Raman spectra of such isotopically labelled bilayer graphene samples, before and after oxygen plasma treatment. The top layers in 2-LG, labelled with $^{12}C$, are registered at higher frequencies than the bottom layers, labelled with $^{13}C$. As can be seen, the D band is almost imperceptible in as-grown graphene, and it becomes more intense upon exposure to the oxygen plasma. In 2-LG

T, the D band arising from the top layer (D ≈ 1350 cm$^{-1}$) is clearly stronger than the D band arising from the bottom layer (D ≈ 1295 cm$^{-1}$), while in the case of 2-LG AB, both layers exhibit similar D bands and Area(D)/Area(G) ratios, see Table 1.

It must be noted that for oxygen plasma treatment, the defects can be assumed to only be created in the top layer of bilayer graphene, because the bottom layer is protected by the top graphene layer. Strictly speaking, this statement is valid in the low-defect-density regime where only sp$^3$ defects are formed. In contrast, for defects created by high energy ion bombardment, the top layer is not protecting the bottom graphene layer and presumably the same number of defects is created in the top and the bottom layers [33].

It has previously been observed that Bernal-stacked layers are strongly connected, with top and bottom layers behaving similarly, while the interaction between two turbostratic layers is weaker and the layers are almost independent of each other [34]. Our results are in agreement with this principle as they show that in Bernal-stacked graphene the defects induced in the top layer equally affect the bottom layer, while in turbostratic layers the bottom layer is much less affected, as it is more independent from the top layer. To explain the strong intensity of the D band in the presumably undefected bottom graphene layer, we suggest that for 2-LG AB the phonons in the bottom layer can be scattered by the defects in the top layer. It has been shown that phonons from different layers in AB-stacked bilayer graphene can interact in one scattering process [35]. For instance, the 2D mode in graphene, where one phonon can originate in the top layer and the second phonon in the bottom layer. For isotopically labelled 2-LG AB, this effect results in a single broad 2D band, as also shown in Figure 5.

As can be seen in Figure 5 and Table 1, the intensity of the D mode is almost the same for the top and bottom layers in 2-LG AB. Assuming that no defects are created in the bottom layer, the total area of the D band is almost twice as large as it should be for a given number of defects in the top graphene layer. These observations suggest that 2-LG AB behaves like one entity, thus an sp$^3$ defect in one layer is observed in the Raman spectra of both layers. For 2-LG T, an increase of the D band intensity for the bottom graphene layer is also observed. Nevertheless, the intensity of the D band of the bottom layer is only about 20% of the total D band area (top layer + bottom layer). This can be understood by either assuming that there is also some scattering of the bottom layer phonons by top layer defects, which is possible at angles close to Bernal stacking, or by the pres-ence of some larger vacancies in the top layer, which would allow the oxygen plasma to also induce defects in the bottom layer. Note that in the case of 2-LG T, the number of defects in the top layer is much larger than for 2-LG AB, presumably due to the higher reactivity of 2-LG T, see Figure 5 and Table 1.

The scattering of the phonons in the bottom layer by defects in the top graphene layer has important consequences for defect quantification in multilayered graphene samples, because the number of defects is often estimated via the intensity of the D band. Consequently, for a proper comparison between the D/G ratios among studied samples, one has to take into account the different meanings of the D/G ratio in monolayers and bilayers. In both cases, the D/G ratio expresses the proportion of defective sites to the total number of carbon atoms. In a monolayer, this relative value can be easily converted to the total amount of defects using proper formulas [5]. In other words, regarding 1-LG samples the expressions "D/G ratio" and "number of defects" are often interchangeable in the literature. However, in bilayers, the situation is more complex because the intensity of the D mode of one layer is affected by defects in both layers. Note that the defect generation by plasma exposure used here results in defects located in the top layer (at least in low defect density regime), whereas for example high energy ion bombardment yields an even distribution of defects between the top and bottom graphene layer as mentioned above. In the latter case the D modes of both layers would be affected by the defects in both graphene layers. Consequently, when comparing defects on bilayers, "the D/G ratio" and "the total number of defects" are two different parameters and need to be used with care.

In general, for non-labelled 2-LG samples, the D and G modes originate from phonons of both the top and bottom layers; hence it becomes more complex to estimate accurately the number of defects via the Area(D)/Area(G) ratio. Nonetheless, taking into account the results obtained on isotopically labelled samples, we can now correct the D/G and D/2D ratios, and estimate the number of defects for the bilayer samples shown in Figure 2. Assuming that for 2-LG the defects are created only in the top layer, one should increase the number of defects by a factor of two, because the D/G or D/2D ratios reflect the contribution of the G or 2D mode both from the top and the bottom layer. However, for 2-LG AB in Figure 5, we see a strong increase of the D mode intensity of the bottom layer. The D mode intensity of the bottom layer is almost the same as in the top layer. Assuming that there are no defects in the bottom layer, the total area of the D mode (bottom + top layer) should be divided by a factor of two for the purpose of calculation of the number of defects. Consequently, for 2-LG AB these two corrections are almost fully compensated and the 'corrected' D/G or D/2D ratios are similar to the original, shown in Figure 2. On the other hand, approximately 80% of the total area of the D mode for 2-LG T originates from the top layer and the remaining 20% from the bottom layer. Consequently, assuming that the defects are only in the top layer, a factor of 1.6 (0.8 because of the 80% contribution of the top layer D band and 2 because of the contribution of the G band from the bottom layer), needs to be included to calculate correctly the number of defects. Correcting the data of Figure 2f accordingly, we observe that the number of defects for the top layer of 2-LG T and 1-LG are in fact similar (see Figure S4 in the supporting information), which is in agreement with our expectations, because the top layer should be independent of the bottom graphene layer in turbostratic graphene. Hence, this result in turn verifies our previous assumptions on scattering of the bottom layer phonons by the top layer defects.

Finally, in order to demonstrate the practical consequences of our work in the quantification of defects, the number of defects for the sample exposed to 44 s of oxygen plasma was calculated. The density of defects $n_D$ (cm$^{-2}$) in monolayer graphene is simply given by the equation $n_D = 7.3 \times 10^9 \times (I_D/I_G) \times E_L^4$ [5], in which $E_L$ = 2.33 eV in our study (note that in the equation the intensity ratio, $I_D/I_G$, instead of the area ratio, $A_D/A_G$, is used). However, for graphene bilayer the situation becomes more complex, as defects are mostly created in the top layer, but the D/G ratio accounts for both layers. Assuming that the defects are only created in the top graphene layer, only the contribution of the top layer should be considered in the total intensity of the D and G modes used in the $I_D/I_G$ ratio. The contributions of the two layers can be distinguished when isotopic labelled graphene is used (see Figure 5). In this way, $I_D/I_G$ ratio (2-LG AB) = 0.38 and $I_D/I_G$ (2-LG T) = 0.94, which gives a defect density of 0.8 and 2.0 x 10$^{11}$ cm$^{-2}$, respectively. However, for standard non-labelled graphene bilayers, one cannot distinguish the top layer from the bottom layer. In this way, the total intensity of the D mode and only ½ of the intensity of the G mode (assuming that the top and bottom layer contribute equally to the total intensity of the G mode) would be used in the calculation of the $I_D/I_G$ ratio. For comparison, we used these assumptions to estimate the number of defects in the same labelled graphene sample exposed 44 s to the oxygen plasma. We obtained $I_D/I_G$ ratio (2-LG AB) = 0.65 and $I_D/I_G$ (2-LG T) = 1.25, which would result in a defect density of 1.4 and 2.7 x $10^{11}$ cm$^{-2}$, respectively. This example clearly demonstrates that if the phonon scattering in the non-defected bottom layer is not considered, one can strongly overestimate the number of defects in the bilayer graphene samples.

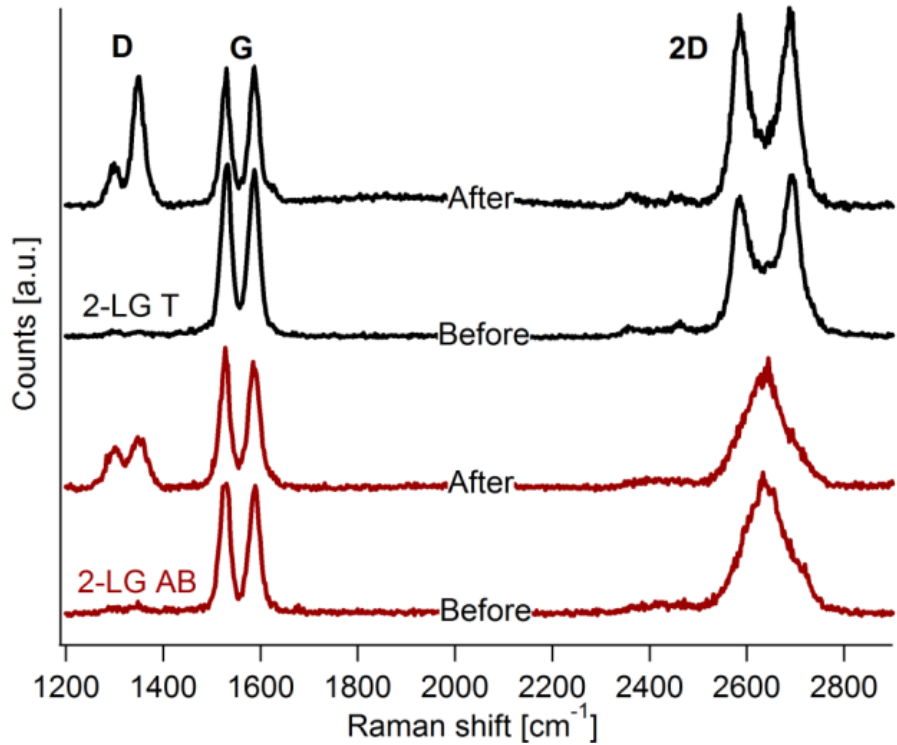

**Figure 5- Raman spectra of isotopically labelled graphene of 2-LG AB and 2-LG T, before (0 s) and after (44 s) oxygen plasma treatment, acquired in the same area of the sample.**

**Table 1- Area(D)/Area(G) ratio of isotopically labelled 2-LG AB and 2-LG T, before (0 s) and after (44 s) oxygen plasma treatment, in the same area of the sample.**

|         | Time of exposure (s) | Area(D)/Area(G) Top layer | Area(D)/Area(G) Bottom layer |
|---------|----------------------|---------------------------|------------------------------|
| 2-LG AB | 0                    | 0.04                      | 0.03                         |
|         | 44                   | 0.50                      | 0.40                         |
| 2-LG T  | 0                    | 0.01                      | 0.02                         |
|         | 44                   | 1.13                      | 0.29                         |

Conclusions

In summary, we have produced defected graphene in a controlled manner, studying its properties through Raman spectroscopy. Our findings show that at similar plasma conditions 1-LG graphene and the top layer of 2-LG T are more susceptible to becoming defected than 2-LG AB. Isotopically labelled samples allowed us to go further and distinguish the effect of the formation of defects on the D mode of the top and bottom layers of bilayer graphene samples. It was observed that bottom layers of graphene bilayers exhibit the D band despite the fact that no defects should be created in the bottom layer. In addition, for 2-LG T, the D mode of the bottom layer is much weaker than in the case of the Bernal-stacked layers. This result was explained by the scattering of phonons of the bottom layer by defects in the top layer of the AB-stacked bilayer. Our findings indicate that the estimation of the number of defects based on the D mode intensity can result in an overestimation of defects by up to a factor of two for 2-LG AB. This result is of high relevance considering the practical importance that Raman spectroscopy has in the quantification of defects in graphene samples.

**Acknowledgments**

The authors acknowledge the support of MSMT project (LH 13022).The authors are grateful to prof. Ado Jorio (UFMG, Belo Horizonte) for the discussion and comments.

**Appendix A. Supplementary data**

Raman D/G ratio for graphene treated at different gas pressures and plasma powers. Optical images of the analysed area before and after plasma treatments. Intensity ratio of D and D' bands as a function of the plasma treatment time. Corrected D/G and D/2D area ratios. Representative histograms for D band FWHM, frequency and area.